\documentclass[lettersize,journal]{IEEEtran}
\usepackage{amsmath,amsfonts}
\usepackage{algorithmic}
\usepackage{algorithm}
\usepackage{array}
\usepackage[caption=false,font=normalsize,labelfont=sf,textfont=sf]{subfig}
\usepackage{textcomp}
\usepackage{stfloats}
\usepackage{url}
\usepackage{verbatim}
\usepackage{graphicx}
\usepackage{cite}
\usepackage{lmodern}
\usepackage{multirow}
\usepackage{comment}
\usepackage{graphicx}
\usepackage{xcolor} 
\usepackage{colortbl}
\usepackage{caption}
\usepackage{amsthm}

\begin{document}

\title{Backward Stochastic Differential
Equations-guided Generative Model for Structural-to-functional Neuroimage Translator}

\author{Zengjing Chen, Lu Wang, Yongkang Lin, Jie Peng, Zhiping Liu,\\
Jie Luo, Bao Wang, Yingchao Liu, Nazim Haouchine, Xu Qiao*}

\maketitle

\markboth{Journal of \LaTeX\ Class Files,~Vol.~14, No.~8, August~2021}%
{Shell \MakeLowercase{\textit{et al.}}: A Sample Article Using IEEEtran.cls for IEEE Journals}


\section*{S1. Methods}

\subsection{Datasets and Patient Cohorts}

The training dataset encompasses two components: the first is the open-source UCSF-PDGM dataset, while the second is derived from MRI scans focused on brain tumors, conducted between December 2016 and March 2020 in the Department of Radiology at Shandong Provincial Hospital, China. These scans employed SCALE-PWI to produce quantitative CBV maps. The overall generation process for the latter dataset, intended for model training, adheres to a structured timeline (see details in the Methods section). It commences with a T2WI lasting 1 minute and 10 seconds, followed by precontrast-T1WI spanning 50 seconds. Next, a T2-FLAIR imaging is performed for 2 minutes and 30 seconds, succeeded by an ADC mapping that takes 1 minute and 17 seconds. Subsequently, the SCALE-PWI protocol involves three stages: Single-slice T1 Mapping (40 seconds), DSC MRI lasting 60 seconds, and a repeat of Single-slice T1 Mapping. A 46-second delay is observed after the initiation of these three steps, preceding the administration of a contrast agent. Following the SCALE-PWI stages, the protocol proceeds with a postcontrast T1WI lasting 50 seconds, culminating in a 3D T-MPRAGE scan spanning 5 minutes.

The training dataset comprises 505 image sets originating from 256 patients, encompassing multiple test results per individual. The test set, meanwhile, consists of 216 image data sets from 206 patients, specifically including subtypes such as glioblastoma, brain metastasis, radionecrosis, and recurrence for validation purposes. The study was granted approval by the local ethics committee in Shandong Provincial Hospital (Issued No. 2019–272), and all experiments adhered strictly to the principles outlined in the Declaration of Helsinki. Due to the retrospective nature of the study, the requirement for informed consent was waived.

\subsection{The detailed information of MRI protocols}

The details regarding the 721 MRI dataset used for CBV generation are as follows. The inclusion criteria encompassed the confirmation of both primary and recurrent brain tumors through pathological examination results, the validation of brain metastases via pathological examination, and the availability of follow-up MRI examinations for patients fulfilling the aforementioned criteria. Conversely, the exclusion criteria encompassed scenarios where the quality of CBV maps was impaired due to metal-induced susceptibility artifacts, cases where the automatic generation of quantitative CBV maps failed owing to registration errors stemming from patient motion, situations involving compromised image quality of standard MRI examinations, and individuals under the age of 18 years.

All patients underwent imaging in the supine position using a 3 T MRI scanner (Magnetom, Skyra; Siemens Healthineers) with a 20-channel transmit/receive quadrature head-and-neck coil. A standardized imaging protocol was applied to all patients, encompassing axial T2-weighted, pre-contrast T1-weighted, T2-FLAIR sequences, and DWI with b-values of 0 s/mm$^2$ and 1,000 s/mm$^2$. Subsequently, bookend dynamic susceptibility contrast (DSC) perfusion- weighted images were acquired after a 46-second injector delay, following which a bolus of 0$\cdot$2 mmol per kg bodyweight of contrast agent (GdDTPA, Magnevist; Schering) was administered, followed by a 20 ml saline flush. The injection velocity was set as 4$\cdot$0 ml/s (typically exceeding 4.5 ml/s). Following this, axial T1C imaging was performed. Finally, three-dimensional T1-weighted magnetisation-prepared rapid gradient-echo images were acquired. Throughout the scanning process, the slice positions for all imaging sequences remain identical. All 2D MRI sequences were acquired with the same imaging scale, position, and slice thickness, facilitating registration across different modalities. The comprehensive parameters are detailed as follows:

\begin{itemize}

\item For T2-weighted imaging (T2WI): repetition time (TR): 3700 ms, echo time (TE): 109 ms, slice number: 19, field of view (FOV): 220 mm, slice thickness: 5 mm, distance factor: 30\%, flip angle (FA): 150, voxel size: 0.3×0.3×5.0 mm$^3$, accelerate factor: 2, bandwidth: 220 Hz/Px, echo spacing: 9.9 ms.

\item For precontrast and postcontrast T1-weighted imaging (T1C): TR: 1820 ms, TE: 13 ms, slice number: 19, FOV: 230 mm, slice thickness: 5 mm, distance factor: 30\%, FA: 150, inversion time (TI): 825 ms, voxel size: 0.4×0.4×5.0 mm$^3$, accelerate factor: 2, bandwidth: 260 Hz/Px, echo spacing: 13 ms.
\item For T2-weighted and fluid-attenuated inversion recovery imaging (T2F): TR: 8000 ms, TE: 81 ms, slice number: 19, FOV: 220 mm, slice thickness: 5 mm, distance factor: 30\%, FA: 150, inversion time (TI): 2370 ms, voxel size: 0.7×0.7×5.0 mm$^3$, accelerate factor: 2, bandwidth: 289 Hz/Px, echo spacing: 9.02 ms.
\item For Diffusion-weighted imaging (DWI): TR: 3700 ms, TE1: 65 ms, TE2: 104 ms, slice number: 19, FOV: 230 mm, slice thickness: 5 mm, distance factor: 30\%, FA: 180°, voxel size: 1.4×1.4×5.0 mm$^3$, acceleration factor: 2, bandwidth: 919 Hz/Px, echo spacing: 0.36 ms, diffusion directions: 3, diffusion mode: 3-Scan trace, diffusion weighting: 2, noise level: 100, b value: 0 and 1000.
\item For postprocessing of apparent diffusion coefficient (ADC) map: Centralized data analysis was conducted at a designated site to derive the ADC from DWI images, utilizing a monoexponential fit between the acquired b = 0 and b $>$ 0 s/mm$^2$ value pairs. The calculation was executed for three distinct DWI directions to characterize each individual gradient channel. The diffusion gradient direction in magnet coordinates was extracted from the DICOM header, which was allocated to a specific gradient channel. However, the DICOM header did not provide details on DWI image postprocessing, such as spatial filtering, but individual sites confirmed that optional filtering was excluded. To address channel-specific b $>$ 0 s/mm$^2$ image distortion resulting from eddy currents, a two-dimensional full-affine co-registration of the b = 0 s/mm$^2$ image was implemented for
DWI data exhibiting substantially differences in  phantom tube displacements and/or misshaping ($>$5mm) across different gradient channels (DWI directions). For systems with significant initial distortions, the co-registration efficiency was visually assessed by examining the consistency in phantom tube position and shape for all DWI directions within each image slice prior to ADC map generation. The co-registration process effectively mitigated major distortions, thereby the uniformity of the resulting ADC map (reducing histogram width) for systems with high eddy current distortions on the selected gradient channels. The ADC was calculated on a pixel-by-pixel basis.
\end{itemize}
\subsection{Bookend DSC-PWI and quantitative CBV map}

In this study, we employed scale-PWI, a prototype bookend DSC-PWI sequence provided by Siemens Healthineers. The sequence seamlessly integrated pre- and postcontrast T1 mapping into the GRE-EPI sequence for DSC-PWI, while introducing a consistent “gradient noise” between T1 mapping and the DSC-PWI scan to mitigate head motion. The imaging parameters of Scale-PWI were as follows: TR/TE, 1,600 ms/30 ms; bandwidth, 1,748 Hz/pixel; 21 axial slices; field of view, 220 × 220 mm; voxel size, 1$\cdot$8×1$\cdot$8×4 mm$^3$; slice thickness, 4$\cdot$0 mm, and flip angle, 90°. For each slice, 50 measurements were acquired to facilitate the bookend DSC-PWI analysis. The quantification of CBV relied on the bookend technique, where the absolute CBV value was derived from the change in white matter’s signal intensity before and after the administration of the contrast agent.


\subsection{Image Pre-processing}
The pre-processing pipeline for all MR images encompasses three consecutive stages to ensure their prime suitability as inputs for the model:

Step 1 (Alignment): A rigorous registration procedure is executed on T1W1 images, ADC images, and the targeted CBV images, leveraging T1C images as a reference to achieve a unified spatial alignment.

Step 2 (Skull Stripping): Employing the FMRIB Software Library (FSL), the images undergo skull stripping, a crucial step to eliminate non-brain structures, thereby refining the focus on relevant neurological information.

Step 3 (Data Augmentation): Tailored data augmentation techniques are systematically applied, with a particular emphasis on the brain region. This step aims to bolster the diversity and robustness of the dataset, ultimately contributing to the enhanced generalization and overall performance of the model.

\subsection{BGM Structure}
Consider the following FBSDE:
\begin{equation}
	\begin{aligned}
		X_t &= \zeta + \int_{0}^{t} b(s,X_s)ds + \int_{0}^{t} \sigma(s,X_s)dW_s \\
		Y_t &= \xi + \int_{t}^{T} f(s,X_s,Y_s,Z_s)ds - \int_{t}^{T}Z_sdW_s
	\end{aligned}
\end{equation}
Here, $X_0 = \zeta$ represents the input to the model, sampled from multi-modal sample data, and the terminal value $Y_T = \xi$ is assumed to adhere to follow the data distribution of the target image.

Utilizing the Feynman-Kac formula, we can establish a connection between the FBSDE and the PDE. Let $Y_t = u(t,X_t)$, it follows that $u(t,x)$ satisfies the corresponding PDE. This also validates the existence of a reasonable mapping $u$ between $Y_t$ and $X_t$. Therefore, to obtain the output from multimodal sample input $\zeta$ to the target CBV image $\xi$, the aforementioned FBSDE provides a sufficient evolution process from $\zeta$ to $\xi$. Next, we delve into the implementation of these equations. Given a partition of the time interval $[0,T]$: $0 = t_0 < t_1 < \cdots < t_N = T$, employing  the Euler forward discrete scheme for both the forward and backward processes, we arrive at:
\begin{equation}
	\begin{aligned}
		X_t &\approx X_{t_n} + b(t_n,X_{t_n})\Delta t_n + \sigma(t_n,X_{t_n})\Delta W_{t_n} \\
		Y_t &\approx Y_{t_n} - f(t_n,X_{t_n},Y_{t_n},Z_{t_n})\Delta t_n + Z_{t_n}\Delta W_{t_n}
	\end{aligned}
\end{equation}
where $\Delta t_n = t_{n+1} - t_n$ and $\Delta W_n = W_{n+1} - W_n$.
\par
\begin{figure*}[http]
    \centering
    \includegraphics[width=.8\linewidth]{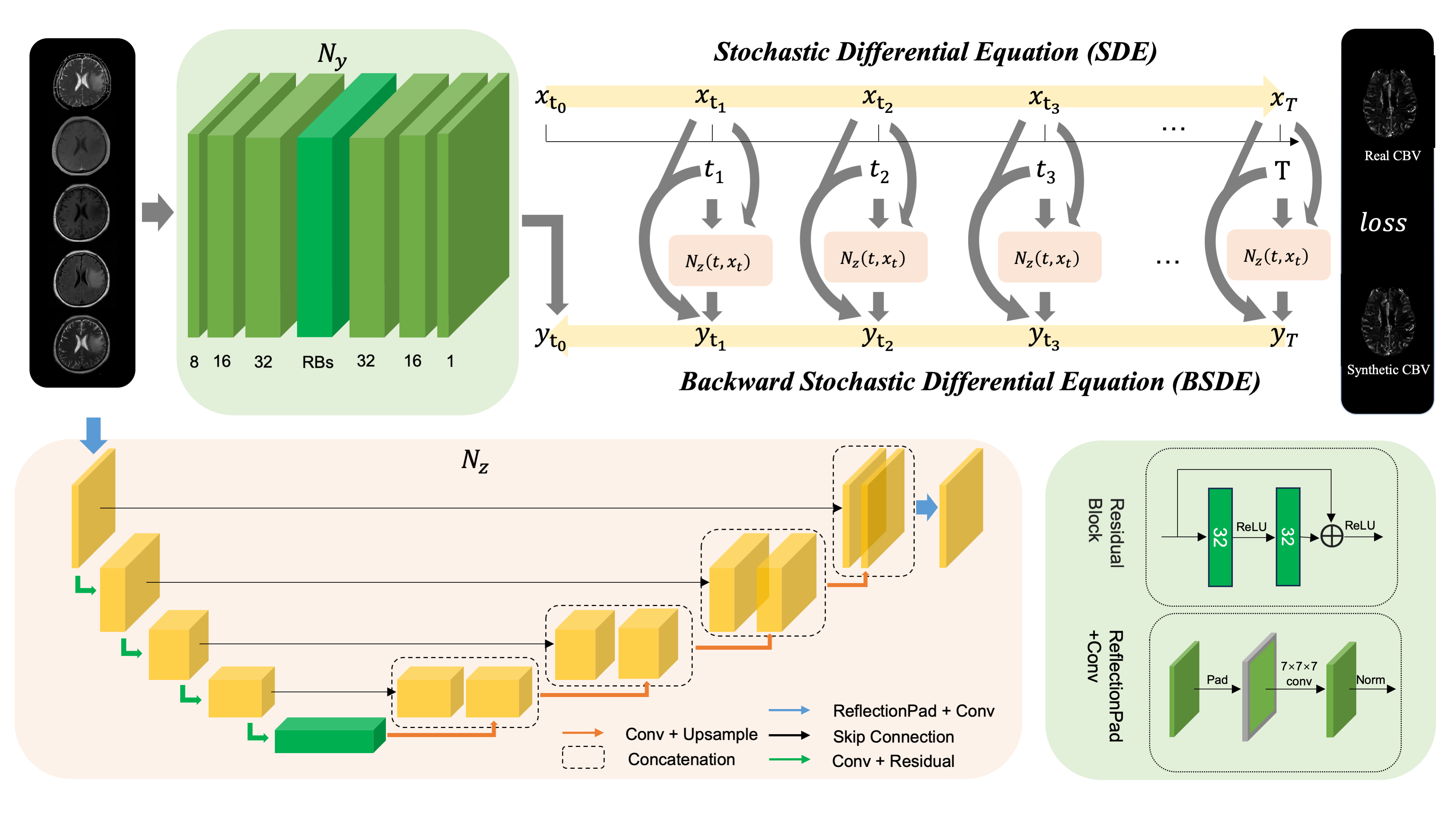}
    \caption{\textbf{Flowchart for synthesizing cerebral blood volume (CBV) maps from standard MRI sequences using BGM.}}
    \label{fig_network}
\end{figure*}

Based on this framework, with the knowledge of the drift function $b(t,x)$ and diffusion function $\sigma(t,x)$ of the forward process, as well as the generator function $f(t,x,y,z)$ of the BSDE, we only need $Y_0$ and the control process $Z_t$ to comprehensively understand the solution $Y_{t}$ of the FBSDE. Furthermore, it is noteworthy that the computation of the BSDE is intricately intertwined with its corresponding generator. As depicted in the arrow flow diagram in Fig. \ref{fig_network}, $Y_{t}$ is related to $s$ and $N_z(t,X_t)$, which aligns with the generator in the BSDE equation.

\section*{S2. Necessary Mathematical}

\subsection{Notations}

Let $(W_t)_{t\geq0}$ be a standard one dimensional Brownian motion
on a complete probability space $(\Omega,\mathcal{F},P)$. By
$\mathbb{F}=\{\mathcal{F}_t,0\leq t\leq T\}$, we denote the natural
filtration generated by $\{W_t\}_{0\leq t\leq T}$ and augmented by
all $P$-null sets. Fixed a time $T>0$, we shall introduce the following spaces:
\begin{itemize}
\small{
\setlength{\itemsep}{0pt}
    \item $M\left([0,T];\mathbb{R}^{n}\right)$: the set of $\mathbb{R}^{n}$-valued progressively measurable process $\phi$ satisfying
  $\mathbb{E}\left[\int_{0}^{\mathrm{T}}\lvert \phi_{t}\rvert d t\right]<\infty$.
    \item $C_b^{1,2}([0,T) \times \Gamma^{n})$: the space of functions $u(t,x)$ on $[0,T) \times \Gamma^{n}$, which are bounded continuously differentiable in $t$ and twice continuously differentiable in $x$.}
\end{itemize}

\subsection{Backward Stochastic Differential Equations}
We provide a brief review of BSDE and the necessary theories used in this work. Consider the following classical structure of BSDE:
\begin{equation}
    \label{BSDE_formal}
\left\{\begin{array}{l}
				-dY_t=g(t,Y_t,Z_t)dt-Z_tdW_t,\\
				Y(T)=\xi,
			\end{array}\right.
\end{equation}
or equivalently,
$$
Y_t = \xi + \int_t^T g(s,Y_s,Z_s)ds - \int_t^TZ_sdW_s
$$
where $\xi$ is a given $\mathcal{F}_T$-measurable random variable, $g(t,y,z)$ is a deterministic real function for all $(t,y,z)\in [0,T]\times \mathbb{R} \times \mathbb{R}$, and the process $(Y_t,Z_t)$ is the solution. Let us give the following assumptions:
\begin{itemize}
\small{
\item \textbf{(A1)} 
 $g:\Omega\times[0,T]\times\mathbb{R}^n\times\mathbb{R}^{n\times d}\rightarrow\mathbb{R}^n$ and for any $(y,z)\in\mathbb{R}^n\times\mathbb{R}^{n\times d}$, $g(\cdot,
        y,z)$ is a $\mathbb{R}^n$-valued ~$\mathcal{F}_t$-adapted process, it satisfies
        $$\int^T_0\lvert g(\cdot,0,0)\rvert ds\in L^2(\Omega,\mathcal{F}_T,P;\mathbb{R}^n).$$
\item \textbf{(A2)}     
        $g$ satisfies the Lipschitz condition: there exists a constant $C>0$, such that $\forall y,y'\in\mathbb{R}^n,\,\,z,z'\in\mathbb{R}^{n\times d}$,
        $$\lvert g(t,y,z)-g(t,y',z')\rvert \leq C(\lvert y-y'\rvert+\lvert z-z'\rvert).$$}
\end{itemize}
Based on the above assumption, we present the following theorem regarding the existence and uniqueness of BSDE in \cite{A1982}:    

\textbf{Theorem 1:}
Let the generator $g$ satisfy the assumptions $\textbf{(A1)}-\textbf{(A2)}$, for any given terminal condition $\xi\in L^2(\Omega,\mathcal{F}_T,P;\mathbb{R}^n),$ BSDE \eqref{BSDE_formal} admits a unique solution, i.e., there exists a unique $\mathcal{F}_t$-adapted process $(Y_{\cdot},Z_{\cdot})\in\mathcal{M}(0,T;\mathbb{R}^n\times\mathbb{R}^{n\times d})$ that satisfies \eqref{BSDE_formal}.

The following is the nonlinear Feynman-Kac formula, which establishes the connection between PDEs and BSDEs. It serves as the theoretical foundation for the network setup for $Y$ and $Z$ in this work. 
For any given $(x,t)\in\mathbb{R}^n\times[0,T]$, let the process $(X^{x,t}_s){t\leq s\leq T}$ satisfy the following SDE:
\begin{equation}
\label{8.17}
\quad\left\{\begin{array}{l}
    dX^{x,t}_s=b(X^{x,t}_s)ds+\sigma(X^{x,t}_s)dB_s,\\
    X^{x,t}_t=x.\\
\end{array}\right.
\end{equation}
where $b(x):\mathbb{R}^n\rightarrow\mathbb{R}^n,\,\sigma(x):\mathbb{R}^n\rightarrow\mathbb{R}^{n\times d}$ are given functions satisfying the Lipschitz condition and linear growth condition.

\textbf{Theorem 2:}
Considering the process $X^{x,t}_s$ in \eqref{8.17}, we introduce the coupled BSDE:
\begin{equation}
\label{8.22}
\left\{\begin{array}{l}
            dY^{x,t}_s=f(X^{x,t}_s,Y^{x,t}_s,Z^{x,t}_s)ds-Z^{x,t}_sdB_s,\\
            Y^{x,t}_T=\Phi(X^{x,t}_T).
        \end{array}\right.
\end{equation}
 If the function $u(x,t)$ defined as follows:
\begin{equation}
\label{8.20}
u(x,t)=Y^{x,t}_s  \vert _{s=t}:\mathbb{R}^n\times[0,T]\rightarrow\mathbb{R}^m
\end{equation}
belongs to $C^{2,1}_b$, then it is also the unique $C^{2,1}_b$ solution of the PDE
\begin{equation}
\label{8.23}
\left\{\begin{array}{l}
            \frac{\partial u}{\partial t}+\mathcal{L}u+f(x,u,u_x\,\sigma)=0,\\
            u(x,T)=\Phi(x),\\
        \end{array}\right.
\end{equation}
where $\mathcal{L}$ second-order elliptic operator:
			$$\mathcal{L}u=\frac{1}{2}\sum^m_{i,j}[\sigma\sigma^{\tau}]_{i,j}(x)\frac{\partial^2 u}{\partial x_i \partial x_j}
			+\sum^m_{i}b_i(x)\frac{\partial u}{\partial x_i}.$$
Conversely, if PDE \eqref{8.23} has an $C^{2,1}_b$ solution, then the solution is unique and the equation \eqref{8.20} holds.

\section*{S3. Derivation of the BSDE with K-ignorance}

Consider the following control problem with the state given by the following SDE
\begin{equation}
\label{sde1}
\left\{\begin{aligned}
&dX_s=u_s\,ds+\,dW_s,\qquad,\\ &X_0=x,\; s\in[0,T],
\end{aligned}\right.
\end{equation}
where the admissible control satisfies
$u_s: [t,T]\rightarrow [-k,k]$ with $k>0$ is any give positive number.
The cost functional is defined as follows
\begin{equation}\label{cost}
J(u)=E[\varphi(x_T)],
\end{equation}
where $\varphi:\mathbb R\rightarrow\mathbb R$ is a symmetric function.
An admissible
control $u^*_s\in\mathcal U$ is called optimal if $J(u^*_s)=\inf_{u_s\in \mathcal{U}} J(u_s)$.

Next we give a representation of the cost function which relates the above control problem to a kind of nonlinear BSDE. Specifically, we will show that
$
\inf _{u_s \in \mathcal{U}} E [\varphi (x_T)]=y_0,
$
where $ (y_s, z_s)$ is the solution of following BSDE
$$
y_s=\varphi (x+\sigma W_{T})-k \int_s^{T} \lvert z_r \rvert d r-\int_s^{T} z_r d W_r .
$$
 Above all, note that
$$
\begin{aligned}
 \inf _{u_s \in \mathcal{U}} E [\varphi (x_T ) ]
& =  \inf _{u_s \in \mathcal{U}} E [\varphi (x+\int_t^T u_s d s+ W_T- W_t ) ] \\
& =  \inf _{\widetilde{u}_s \in \widetilde{\mathcal{U}}} E [\varphi (x+\int_0^{T-t} \tilde{u}_s d s+ W_{T-t} ) ],
\end{aligned}
$$
where $\tilde{\mathcal{U}}$ is the set of $ \{\mathcal{F}_s \}$-adapted stochastic processes $\tilde{u}_s$ such that $\lvert \tilde{u}_s \rvert \leq k, 0 \leq s \leq T$.

First, define $a_r=k \operatorname{sgn} (z_r ), \tilde{W}_s=W_s+\int_0^s a_r d r$, then $\tilde{W}_s$ is a Brownian motion under $Q$, where 
$$ \frac{d Q}{d P} \lvert _{\mathcal F_S}=\exp  (-\int_0^s a_r d W_r- \frac{1}{2} \int_0^s a_r^2 d r ).$$ Therefore,
$$
y_0=\varphi (x+\sigma \widetilde{W}_{T}-\sigma \int_0^{T} a_r d r )-\int_0^{T} z_r d \widetilde{W}_r.
$$
Hence
$$
\begin{aligned}
y_0   & =E_Q [\varphi (x+ \tilde{W}_{T}- \int_0^{T} a_r d r ) ] \\
 & \geq  \inf _{\widetilde{u}_s \in \tilde{\mathcal{U}}} E_Q [\varphi (x+ \widetilde{W}_{T}+\int_0^{T} \tilde{u}_r d r ) ] .
\end{aligned}
$$

Next, for any $ \theta_s \in \tilde{\mathcal{U}}$, consider the following BSDE
$$
y_s^\theta=\varphi (x+ W_{T} )+\int_s^{T} \theta_r z_r^\theta d r-\int_s^{T} z_r^\theta d W_r .
$$

Define $W_s^\theta=W_s-\int_0^s \theta_r d r$, then $W_s^\theta$ is a Brownian motion under $P^\theta$, where $\frac{d p^\theta}{d P} \lvert _{\mathcal{F}_s}=\exp  (\int_0^s \theta_r d W_r-\frac{1}{2} \int_0^s \theta_r^2 d r )$. Therefore, $y_0^\theta=E_{p^\theta} [\varphi (x+\sigma W_{T}^\theta+\sigma \int_0^{T} \theta_{\mathrm{r}} d r ) ].$

By the comparison theorem of BSDE, we have $y_0 \leq y_0^\theta.$ Therefore,
$$
y_0 \leq \inf _{\widetilde{u}_s \in \tilde{\mathcal{U}}} E_Q [\varphi (x+ \tilde{W}_{T}+\int_0^{T} \tilde{u}_r d r ) ] .
$$

Therefore, we have
$$
\inf _{\widetilde{u}_s \in \tilde{\mathcal{U}}} E [\varphi (x+\sigma W_{T-t}+\int_0^{T-t} \tilde{u}_r d r ) ]=y_0 .
$$

Finally, the corresponding Hamiltonian system of the above control problem is the following FBSDE
\begin{equation}\left\{\begin{aligned}
&dx_s=-k sgn(\varphi'(x_s))ds+dW_s,\\
&dy_s=-k\lvert z_s\rvert ds-z_sdW_s,\\
&x_0=x,\ y_T=\varphi(x+W_T).
\end{aligned}\right.\end{equation}

\section*{S4. Analytic posterior given boundary pair}

\textbf{Proposition:}
The posterior of the SDE
\begin{equation}
\left\{\begin{aligned}
&dX_s = -kds + \sigma dW_s,\\
&X_0 = x_0,
\end{aligned}\right.
\end{equation}
given some boundary pair $(X_0,X_1)$ admits an analytic form:
\begin{equation}
\begin{aligned}
p(x_t\lvert x_1,x_0) &= \mathcal{N}(X_t
;\mu_t(X_0,X_1), \Sigma_t),
\end{aligned}
\end{equation}
where 
\begin{equation}
\begin{aligned}
&\mu_t = \frac{\sigma^2_2}{\sigma^2_1 + \sigma^2_2}\mu_1 + \frac{\sigma^2_1}{\sigma^2_1 + \sigma^2_2} \mu_2, \\
& \mu_1 = x_0 -kt,\\
& \mu_2 = x_1 - kt + k, \\
& \sigma_1 = \sigma^2t, \\
& \sigma_2 = \sigma^2 -  \sigma^2t 
\end{aligned}
\end{equation}

\textbf{Proof:}
Applying the $ It\hat{o} $'s formula, we have
\begin{equation}
\begin{aligned}
& dX^2_s = (-2kX_s + \sigma^2) ds + 2\sigma X_sdW_s,\\
& X^2_0 = x,\; s\in[0,T].
\end{aligned}
\end{equation}
Then, we could have different expressions by integrating from $[0,t]$ and $[t,1]$,

\begin{equation*}
\left\{\begin{aligned}
& X_t = x_0 - kt + \int_0^t \sigma  dW_s,\\
& X_t = x_1-kt + k - \int_t^1\sigma dW_s,\\
& X^2_t = X_0^2 + \int_0^t (-2kX_s + \sigma^2) ds + \int_0^t 2\sigma X_sdW_s,\\
& X^2_t = X_1^2 - \int_t^1 (-2kX_s + \sigma^2) ds - \int_t^1 2\sigma X_sdW_s.
\end{aligned}\right.
\end{equation*}
Notice that the expressions of $X^2_t$ still include the term regrading with $X_s$. In order to get rid of this, we introduce the following process
$$ L_t=\exp(-\frac{k^2}{2\sigma^2}t+\frac{k}{\sigma}W_t) $$
Applying $ It\hat{o} $’s formula to $ X^2_tL_t $, we obtain
$$dX_{t}^{2}L_t = \sigma^{2}L_tdt+(2\sigma X_{t}L_{t}+\frac{k}{\sigma}X_{t}^{2}Lt)dW_{t}.$$
Expressed in integral form and taking expectations on both sides, we get
\begin{align*}
\mathbb{E}[X_{t}^{2}]= &\mathbb{E}[\frac{1}{L_t}X_{0}^{2}]+\mathbb{E}\int_{0}^{t}\frac{L_s}{L_t}\sigma^{2}ds\\
&+\mathbb{E}[\frac{1}{L_t}\int_{0}^{t}(2\sigma X_{s}L_s+\frac{b}{\sigma}X_{s}^{2}Ls)dWs]
\end{align*}
Next, let’s calculate the three parts of the integral on the right-hand side:

Part I :
$$\mathbb{E}[\frac{1}{L_t}X_0^2]=X_0^2\mathbb{E}[e^{\frac{k^{2}}{2\sigma^2}t-\frac{k}{\sigma}W_t}]=X_0^2e^{\frac{k^{2}}{\sigma^{2}}t}.$$
Part II :
\begin{equation}
\begin{aligned}
\mathbb{E}\int_{0}^{t}\frac{L_s}{L_t}\sigma^{2}ds&=\mathbb{E}\int_{0}^{t}e^{\frac{1}{2}\frac{k^{2}}{\sigma^2}(t-s)+\frac{k}{\sigma}(W_s-W_t)}\sigma^{2}ds\\
&=\int_{0}^{t}e^{\frac{k^{2}}{\sigma^{2}}(t-s)}\sigma^{2}ds\\
&=-\frac{\sigma^{4}}{k^{2}}e^{\frac{k^{2}}{\sigma^{2}}(t-s)}\vert _{0}^{t}=-\frac{\sigma^{4}}{k^{2}}+\frac{\sigma^{4}}{k^{2}}e^{\frac{k^{2}}{\sigma^{2}}t}
\end{aligned}
\end{equation}
Part III:\\
Next, we will calculate the expectation of 
$$ \frac{1}{L_t}\int_{0}^{t}(2\sigma X_{s}L_s+\frac{k}{\sigma}X_{s}^{2}Ls)dWs $$ 
Define
 $$f_{t}=\frac{1}{L_{t}},g_{t}=\int_{0}^{t}(2\sigma X_{s}L_{s}+\frac{k}{\sigma}X_{s}^{2}L_{s})dW_{s}.$$
Applying $ It\hat{o} $’s formula to $ f_tg_t $, we get
\begin{align*}
d(f_{t}g_{t})=&(\frac{k^{2}}{\sigma^{2}}f_{t}g_{t}-2kX_{t}-\frac{k^{2}}{\sigma^{2}}X_{t}^{2})dt\\
&+(2\sigma X_{t}+\frac{k}{\sigma}X_{t}^{2}-\frac{k}{\sigma}f_{t}g_{t})dW_{t}.
\end{align*}
Integrating from 0 to t, taking expectations,
\begin{align*}
&\mathbb{E}[f_{t}g_{t}]  \\
=&\mathbb{E}\int_{0}^{t}(\frac{k^{2}}{\sigma^{2}}f_{s}g_{s}-2kX_{s}-\frac{k^{2}}{\sigma^{2}}X_{s}^{2})ds \\
=&\int_{0}^{t}(\frac{k^{2}}{\sigma^{2}}Ef_{s}g_{s}-2k(x_{0}-ks)-\frac{k^{2}}{\sigma^{2}}\mathbb{E}[X_{s}^{2}])ds.
\end{align*}
Denote $ F(t)=\mathbb{E}[f_tg_t] $, which is a deterministic function. Taking its derivative, we have
$$ F^{\prime}(t)=\frac{k^{2}}{\sigma^{2}}F(t)-2k(x_{0}-kt)-\frac{k^{2}}{\sigma^{2}}\mathbb{E}[X_{t}^{2}] .$$
This is an ODE, and it's solution is
\begin{align*}
&F(t)=\mathbb{E}[\frac{1}{L_{t}}\int_{0}^{t}(2\sigma X_{s}L_{s}+\frac{k}{\sigma}X_{s}^{2}L_{s})dW_{s}]\\
=&e^{\frac{k^{2}}{\sigma^{2}}t}\int_{0}^{t}e^{-\frac{k^{2}}{\sigma^{2}}s}[-2k(x_{0}-ks)-\mathbb{E}[X_{s}^{2}]\cdot\frac{k^{2}}{\sigma^{2}}]ds\\
=&-(\frac{2\sigma^{2}}{k}x_{0}-2\sigma^{2}t-\frac{2\sigma^{4}}{k^{2}})+(\frac{2\sigma^{4}}{k^{2}}-\frac{2\sigma^{2}}{k^{2}}x_{0})e^{\frac{k^{2}}{\sigma^{2}}t}\\
&-\int_{0}^{t}e^{\frac{k^{2}}{\sigma^{2}}(t-s)}\frac{k^{2}}{\sigma^{2}}\mathbb{E}[X_{s}^{2}]ds,
\end{align*}
then
\begin{align*}
&\mathbb{E}X_{t}^{2}
=e^{\frac{k^{2}}{\sigma^{2}}t}x_{0}^{2}-\frac{\sigma^{4}}{k^{2}}+\frac{\sigma^{4}}{k^{2}}e^{\frac{k^{2}}{\sigma^{2}}t}+(\frac{2\sigma^{2}}{k}x_{0}-2\sigma^{2}t-\frac{2\sigma^{4}}{k^{2}})\\
&+(\frac{2\sigma^{4}}{k^{2}}-\frac{2\sigma^{2}}{k}x_{0})e^{\frac{k^{2}}{\sigma^{2}}t}-e^{\frac{k^{2}}{\sigma^{2}}t}\int_{0}^{t}e^{-\frac{k^{2}}{\sigma^{2}}s}\mathbb{E}[X_{s}^{2}]\cdot\frac{k^{2}}{\sigma^{2}}ds\\
=&e^{\frac{k^{2}}{\sigma^{2}}t}(x_{0}^{2}+\frac{\sigma^{4}}{k^{2}}+\frac{2\sigma^{4}}{k^{2}}-\frac{2\sigma^{2}}{k}x_{0})+(\frac{2\sigma^{2}}{k}x_{0}-2\sigma^{2}t\\
&-\frac{2\sigma^{4}}{k^{2}}-\frac{\sigma^{4}}{k^{2}})-e^{\frac{k^{2}}{\sigma^{2}}t}\int_{0}^{t}e^{-\frac{k^{2}}{\sigma^{2}}s}\mathbb{E}[X_{s}^{2}]\cdot\frac{k^{2}}{\sigma^{2}}ds.
\end{align*}
Let $ G(t)=\mathbb{E}[X_{t}^{2}]\cdot e^{-\frac{k^{2}}{\sigma^{2}}t}, $ then
\begin{align*}
G(t)=&x_{0}^{2}+\frac{3\sigma^{4}}{k^{2}}-\frac{2\sigma^{2}}{k^{2}}x_{0}+(\frac{2\sigma^{2}}{k}x_{0}-2\sigma^{2}t\\
&-\frac{3\sigma^{4}}{k^{2}})e^{-\frac{k^{2}}{\sigma^{2}}t}-\int_{0}^{t}G(s)\cdot\frac{k^{2}}{\sigma^{2}}ds .
\end{align*}
Differentiating with respect to t, we have
\begin{align*}
G^{\prime}(t)
&=(-2kx_{0}+2k^{2}t+\sigma^{2})e^{-\frac{k^{2}}{\sigma^{2}}t}-G(t)\cdot\frac{k^{2}}{\sigma^{2}}.
\end{align*}
It's also an ODE, and its solution is
 $$ G(t)=e^{-\frac{k^2}{\sigma^2}t}[(x_{0}-kt)^{2}+\sigma^{2}t].  $$
then
$$\mathbb{E}[X_{t}^{2}]=G(t)\cdot e^{\frac{k^{2}}{\sigma^{2}}t}=(x_{0}-kt)^{2}+\sigma^{2}t . $$
By direct calculations, we have
\begin{equation*}
\begin{aligned}
&\mathbb{E}[X_t] = X_0 -kt,\\
&\mathbb{E}[X^2_t] - (\mathbb{E}[X_t])^2 = \sigma^2t.
\end{aligned}
\end{equation*}
With similar calculation, we could have the formulation of $\mu_2$ and $\sigma_2$. $\qed$

\onecolumn 
\newpage
\section*{S5. Performance metrics for BGM, GAN, Pix2Pix, and Encoder-Decoder Models.}

\begin{table}[http]
\centering
\captionsetup{labelformat=empty}
\footnotesize{
\caption{\textbf{Table S1.} Performance metrics for BGM, GAN, En-De, Pix2Pix Models in CBV generation task.}
\renewcommand{\arraystretch}{1.5}
\begin{tabular}{ccccc}
\hline
\rowcolor{gray!20}
Methods & Dataset & \multicolumn{1}{c}{LPIPS} & \multicolumn{1}{c}{MAE} & \multicolumn{1}{c}{MSE} \\ \hline
\multirow{2}{*}{BGM} & GBM & 0.2257(±0.0328) & 0.0854 (±0.0271) & 0.0757 (±0.0596) \\ 
& MT  & \textbf{0.2140(±0.0256)} & 0.0859 (±0.0291)  & \textbf{0.0612 (±0.0395)}\\ \hline
\multirow{2}{*}{En-De} & GBM & 0.2362(±0.0292)& 0.1292(±0.0200) & 0.1576(±0.0659) 
\\ & MT  & 0.2269(±0.0215) & 0.1222(±0.0174) & 0.1247(±0.0386) \\ \hline
\multirow{2}{*}{GAN} & GBM & 0.2495(±0.0369)& 0.0865(±0.0164) & 0.0898(±0.0483)  \\
 & MT  & 0.2372 (±0.0294)& \textbf{0.0834(±0.0177)} & 0.0694(±0.0317) \\ \hline
 \multirow{2}{*}{Pix2Pix} & GBM & 0.4390 (±0.0810)& 0.1644 (±0.0410 ) & 0.0697 (±0.0213 )  \\
 & MT &0.4052(±0.0647) &0.1940 (±0.0315) & 0.0841 (±0.0188 ) \\ \hline
\end{tabular}

\vspace{2em}
\begin{tabular}{ccccc}
\hline
\rowcolor{gray!20}
Methods & Dataset & \multicolumn{1}{c}{NCC} & \multicolumn{1}{c}{PSNR} & \multicolumn{1}{c}{SSIM} \\ \hline
\multirow{2}{*}{BGM} & GBM  & \textbf{0.8692 (±0.0774)} & \textbf{31.3380(±2.8176)} & \textbf{0.9147 (±0.0423)}  \\ 
& MT   & 0.8629 (± 0.0818) & 30.4775 (±2.9431) & 0.9119 (±0.0428) \\ \hline
\multirow{2}{*}{En-De} & GBM  & 0.5267(±0.0820)  & 28.2253(±2.9285)  & 0.8310(±0.0456)
\\ & MT  & 0.5819(±0.0573) &  27.3093(±2.5945)  & 0.8380(±0.0417) \\ \hline
\multirow{2}{*}{GAN} & GBM  & 0.7370(±0.0586) &29.9704(±2.5572) & 0.8677(±0.0483) \\
 & MT  & 0.7719(±0.0755) & 29.0018(±2.7520) & 0.8758(±0.0413)\\ \hline
 \multirow{2}{*}{Pix2Pix} & GBM  & 0.3283 (±0.1428 ) &21.9855 (±1.8499 ) & 0.7509 (±0.0533 ) \\
 & MT & 0.4253 (±0.1182 ) & 21.5420 (±2.0930 ) & 0.7423 (±0.0585 )\\ \hline
\end{tabular}
}
\end{table}

\begin{table}[http]
\centering
\captionsetup{labelformat=empty}
\footnotesize{
\caption{\textbf{Table S2.} Performance metrics for BGM, GAN, En-De, CycleGAN, Pix2Pix and DDPM Models in DWI generation task.}
\renewcommand{\arraystretch}{1.5}
\begin{tabular}{cccc}
\hline
\rowcolor{gray!20}
Methods & \multicolumn{1}{c}{LPIPS} & \multicolumn{1}{c}{MAE} & \multicolumn{1}{c}{MSE}  \\ \hline
\multirow{1}{*}{BGM} & \textbf{0.0624(±0.0238)} & \textbf{0.0119(±0.0029)} & \textbf{0.0017(±0.0008)} \\
 \hline
\multirow{1}{*}{En-De} & 0.0631(±0.0185)
  & 0.0125(±0.0028)  & 0.0018(±0.0008) \\
 \hline
 \multirow{1}{*}{GAN} & 0.1221(±0.0225) & 0.0136(±0.0029) & 0.0021(±0.0009) 
 \\ \hline
 \multirow{1}{*}{CycleGAN} & 0.5565(±0.0967) & 0.1554(±0.1554) & 0.0767 (±0.0342)
 \\ \hline
  \multirow{1}{*}{Pix2Pix} & 0.1368(±0.0490) & 0.1012(±0.0898) & 0.0268(±0.0589)
 \\ \hline
  \multirow{1}{*}{DDPM} & --- 
 & 0.2934(±0.2527)
& 0.1821(±0.1475)   
 \\ \hline
\end{tabular}

\vspace{2em}
\begin{tabular}{cccc}
\hline
\rowcolor{gray!20}
Methods & \multicolumn{1}{c}{NCC} & \multicolumn{1}{c}{PSNR} & \multicolumn{1}{c}{SSIM} \\ \hline
\multirow{1}{*}{BGM}  & \textbf{0.9833(±0.0065)} & 36.6243(±2.0939) & \textbf{0.9736(±0.0105)}\\
 \hline
\multirow{1}{*}{En-De}  &0.9826(±0.0065) &  36.3682(±2.0934) & 0.9690(±0.0118)\\
 \hline
 \multirow{1}{*}{GAN}  & 0.9800(±0.0066)  & 35.6079(±1.7898) & 0.9625(±0.0122)
 \\ \hline
 \multirow{1}{*}{CycleGAN} & 0.5638(±0.1768) & 19.8859(±3.1742)  & 0.8421(±0.0882)
 \\ \hline
  \multirow{1}{*}{Pix2Pix} & 0.9842(±0.0098) & \textbf{36.9772(±2.7595)}  & 0.9628(±0.0822)
 \\ \hline
  \multirow{1}{*}{DDPM}
 & 0.9821(±0.0085)&  35.4620(±2.8928)
& 0.8730(±0.1757)
 \\ \hline
\end{tabular}
}
\end{table}

\newpage
\begin{table*}[http]
\centering
\captionsetup{labelformat=empty}
\footnotesize{
\caption{\textbf{Table S3.} Performance metrics for BGM, GAN, En-De, CycleGAN, Pix2Pix and DDPM Models in T1C generation task.}
\renewcommand{\arraystretch}{1.5}
\begin{tabular}{ccccc}
\hline
\rowcolor{gray!20}
Methods & Dataset & \multicolumn{1}{c}{LPIPS} & \multicolumn{1}{c}{MAE} & \multicolumn{1}{c}{MSE}  \\ \hline
\multirow{2}{*}{BGM} & T1+T2+T2FLAIR  & \textbf{0.0535(±0.0194)} & \textbf{0.0111(±0.0033)} & \textbf{0.0024(±0.002)} 
\\
 & T1+T2   & 0.0578(±0.0200) & 0.0112(±0.0033) & 0.0025(±0.0019) \\ \hline
\multirow{2}{*}{En-De} & T1+T2+T2FLAIR & 0.1127(±0.0242)
 &0.0126(±0.0034)& 0.0033(±0.0025)  \\
 & T1+T2  & 0.1020(±0.0248)&0.0124(±0.0034) & 0.0032(±0.0024)  \\ \hline
 \multirow{2}{*}{GAN} & T1+T2+T2FLAIR & 0.1348(±0.0272)
&0.0154(±0.0034)& 0.0042(±0.0027)  \\
 & T1+T2 & 0.1278(±0.0268) & 0.0193(±0.0046)&  0.005(±0.0027)
 \\ \hline
 \multirow{2}{*}{CycleGAN} & T1+T2+T2FLAIR & 0.3980(±0.0502) & 0.1545(±0.0597) & 0.0740(±0.0299) \\
 & T1+T2 &0.3576(±0.0273)& 0.1338(±0.0742)&  0.0980(±0.0287) 
 \\ \hline
 \multirow{2}{*}{Pix2Pix} & T1+T2+T2FLAIR & 0.3166(±0.0307)
& 0.1399(±0.0483) & 0.0521(±0.0183)  \\
 & T1+T2 &0.3161(±0.0294)& 0.1205(±0.0228)&  0.0402(±0.0094)
 \\ \hline
\end{tabular}

\vspace{2em}
\begin{tabular}{ccccc}
\hline
\rowcolor{gray!20}
Methods & Dataset & \multicolumn{1}{c}{NCC} & \multicolumn{1}{c}{PSNR} & \multicolumn{1}{c}{SSIM} \\ \hline
\multirow{2}{*}{BGM} & T1+T2+T2FLAIR  &  \textbf{0.9757(±0.0139)}
 & \textbf{35.012(±1.8132)} & \textbf{0.9459(±0.0183)}
\\
 & T1+T2   & 0.9745(±0.0138)&  34.7334(±1.6242) & 0.9425(±0.0185)\\ \hline
\multirow{2}{*}{En-De} & T1+T2+T2FLAIR  & 0.9661(±0.0176) & 32.7269(±1.416) & 0.9256(±0.0197) \\
 & T1+T2  & 0.9665(±0.0175) & 32.814(±1.4105)  & 0.9257(±0.0209) \\ \hline
 \multirow{2}{*}{GAN} & T1+T2+T2FLAIR  & 0.956(±0.0205) & 31.4976(±1.2116) & 0.9155(±0.0211) \\
 & T1+T2 & 0.9528(±0.0207) & 30.752(±1.3394)&  0.9091(±0.0212) 
 \\ \hline
 \multirow{2}{*}{CycleGAN} & T1+T2+T2FLAIR  & 0.8445(±0.0506) & 24.1360(±1.7773)  & 0.8399(±0.0421)\\
 & T1+T2 & 0.6028(±0.0940) & 17.9917(±0.4893)&  0.8040(±0.0203) 
 \\ \hline
 \multirow{2}{*}{Pix2Pix} & T1+T2+T2FLAIR  & 0.8619(±0.0573) & 24.8925(±1.0058)  & 0.8741(±0.0221) \\
 & T1+T2  & 0.8390(±0.0651) & 24.0361(±0.9340)&  0.8827(±0.0196) 
 \\ \hline
\end{tabular}
}
\end{table*}

\end{document}